\documentclass[10pt]{article}

\usepackage{hyperref}
\usepackage{verbatim}
\usepackage{amsmath}
\usepackage{setspace}
\usepackage{amsfonts}
\usepackage{amssymb}
\usepackage{bold-extra}
\usepackage{graphicx}
\usepackage{algorithm}
\usepackage{algorithmic}
\usepackage{tikz}
\usetikzlibrary{shapes,shadows,arrows}
\usepackage{caption}
\usepackage{setspace}

\usepackage{epstopdf}

\usepackage{bm}
\usepackage{bm}
\newcommand{\vect}[1]{\boldsymbol{\mathbf{#1}}}
\usepackage{subfigure}
\usepackage{graphicx}
\usepackage{epstopdf}
\usepackage{wrapfig}
\usepackage{float}
\usepackage{enumerate}

\usepackage{color}

\usepackage{amsbsy}
\usepackage{dcolumn}
\usepackage{caption}

\usepackage[margin=.5in]{geometry}
\usepackage{epsfig} 
\usepackage{mathptmx} 
\usepackage{times} 
\usepackage{amsmath} 
\usepackage{amssymb}  
\newcommand{\quotes}[1]{``\textbf{#1}''}

\begin{document}


\title{Spontaneous spinning of a rattleback placed on vibrating platform \\
\begin{small}
Aditya Nanda \\ Puneet Singla \\ M. Amin Karami  \\
University at Buffalo, NY
\end{small}
}

\maketitle
\begin{abstract}
\it{This paper investigates the spin resonance of a rattleback subjected to base oscillations which is able to transduce vibrations into continuous rotary motion and, therefore, is ideal for applications in Energy harvesting and Vibration sensing.  The rattleback is a toy with some curious properties. When placed on a surface with reasonable friction, the rattleback has a preferred direction of spin. If rotated anti to it, longitudinal vibrations are set up and spin direction is reversed.

In this paper, the dynamics of a rattleback placed on a sinusoidally vibrating platform are simulated. We can expect base vibrations to excite the pitch motion of the rattleback, which,  because of the coupling between pitch and spin motion, should cause the rattleback to spin. Results are presented which show that this indeed is the case- the rattleback has a mono-peak spin resonance with respect to base vibrations. 

The dynamic response of the rattleback was found to be composed of two principal frequencies that appeared in the pitch and rolling motions. One of the frequencies was found to have a large coupling with the spin of the rattleback. Spin resonance was found to occur when the base oscillatory frequency was twice the value of the coupled frequency.  A linearized model is developed which can predict the values of the two frequencies accurately and analytical expressions for the same in terms of the parameters of the rattleback have been derived. The analysis, thus, forms an effective and easy method for obtaining the spin resonant frequency of a given rattleback. 
 
 Novel ideas for applications utilizing the phenomenon of spin resonance, for example, an energy harvester composed of a magnetized rattleback surrounded by ferromagnetic walls and a small scale vibration sensor comprising an array of several magnetized rattlebacks, are included. }
\end{abstract}

\section{ INTRODUCTION \label{intro}} 

\textit{\begin{large}
Please see \quotes{Energy harvesting using a Rattleback: Theoretical analysis and simulations of spin resonance} in Journal of Sound and Vibration, 369,   pages={195--208}, (2016) for original publication.   \cite{nanda2016energy}
\end{large}
}

\begin{figure}[htp]
\centering
\includegraphics[width=.9\linewidth]{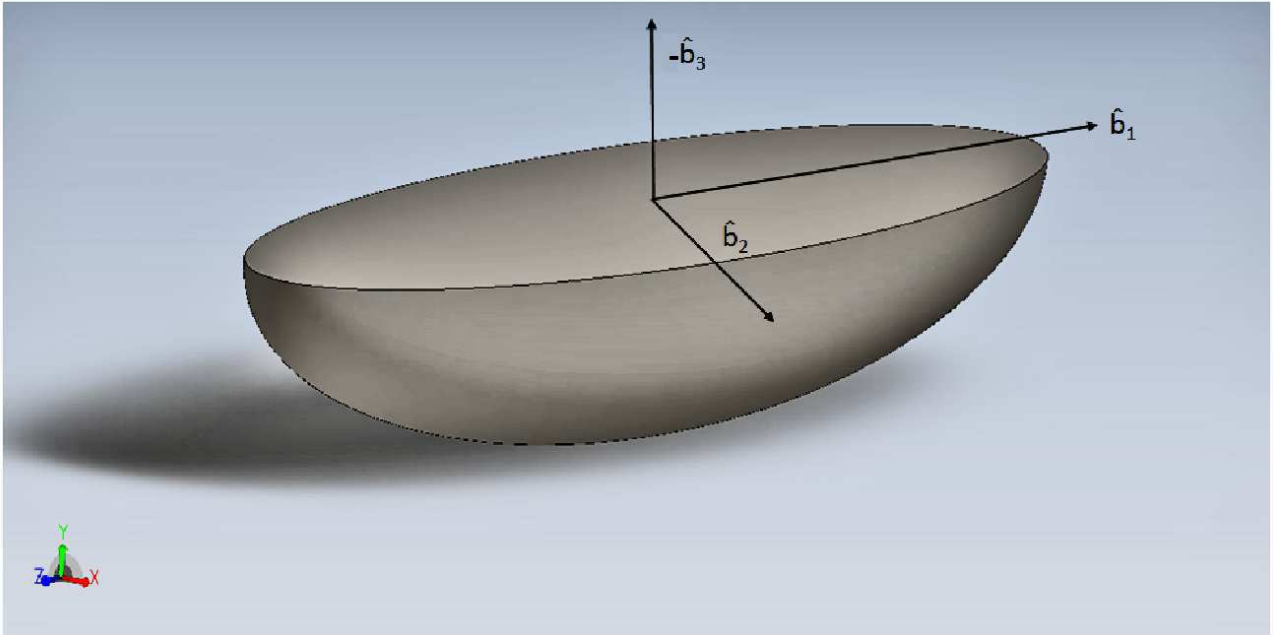} 
\caption{RATTLEBACK}
\label{rattaxes}
\end{figure}

\vspace{.2in}
The rattleback, sometimes also known by the names "celt", "anagyre", "rebellious celt" or "wobblestone", is shaped like a semi-ellipsoidal top and exhibits some very interesting dynamic properties. The most common type of rattleback will rotate about it's vertical axis only in the preferred direction. If spun in the opposite sense, pitch vibrations are set up and the spin reverses. Further, if it is imparted some pitching oscillations, the pitch motion dies down quickly and a spinning motion builds up about the vertical axis along the stable spin direction. Some rattlebacks will reverse when spun in either direction.The rattleback usually has the shape of an semi-ellipsoid as shown in Fig. \ref{rattaxes}. 

The rattleback has been the subject of a large number of analyses since the 1890's  (see \cite{gwalker2,gwalker}). Many numerical analyses(see \cite{boris,garcia,kane}) were done in 1980's after the advent of computing technologies. In Bondi's paper, \cite{bondi}, the author demonstrates that the preference for a spin direction can be readily defined and explained by a small asymmetry between the horizontal principal inertia axes and the directions of maximum curvature(principal axes of curvature). Further, in the paper,  the  equations of motion of a rattleback  are linearized about some nominal spin about the $3-$ axis and a fourth order characteristic equation pertaining to pitching and rolling motion is obtained. A short analysis of the characteristic equation is also done. 

A mathematical treatment of the subject is done by Borisov and Mamaev in \cite{boris}. The authors used a semi ellipsoidal model with offset inertia axes to analyze the presence of strange attractors in the response of an rattleback. In \cite{pascal1983}, the method of averaging is used to integrate the dynamic equation of the rattleback derived by taking into account only first and second order terms while ignoring slipping.  In \cite{markeev1983}, Markeev derives approximate equations of the rattleback by considering small oscillations about the equilibrium. 

Garcia \textit{et al.} in \cite{garcia} carried out numerical integration of the equations of motion  and commented on the behavior of zone 0, zone I and zone II rattlebacks which were first expostulated by Bondi. They further developed a model that incorporated slipping of the point of contact and dissipative forces such as aerodynamic drag forces and contact patch friction.  Zone 0 and Zone I rattlebacks, although similar, differ in the fact that stable spin is not possible in Zone 0 rattlebacks in either direction whereas in Zone I rattlebacks unidirectional stable spin is possible only if energy is higher than a critical value. The rattleback(s) in this paper will be modeled, following \cite{boris}, as a semi ellipsoid with inertia axes slightly offset from the geometric axes. 

In this paper, for the first time in literature, we will demonstrate the existence of spin resonance, analyze the mechanism of it's occurrence and derive a closed form expression for the resonant frequency of a given rattleback in terms of it's inertia parameters. The phenomenon of spin resonance can have novel applications in ambient energy harvesting and vibration sensing since it is fairly straightforward to convert rotations to electric energy via electromagnetic induction. 
  
This paper is organized in six sections. The second section(\ref{sec2}) reviews the non-linear equations of motion governing the motions of a rattleback using Newtonian formulation. Results demonstrating the occurrence of spin resonance with respect to base vibrations are presented in the third section(\ref{sec3}). The fourth section(\ref{sec4}) proposes a mechanism explaining the resonance and we present a linearized model of the rattleback that is able to successfully predict the spin resonant frequency and an expression for the same as a function of mass and inertia parameters of the rattleback is derived. Supporting arguments and validation of this analysis is presented at the end of section four(See subsection \ref{validate}). Section five(\ref{appli}) presents prototypical ideas for a Energy Harvester and a Vibration sensor that utilize the phenomenon of Spin resonance. In the final section(\ref{conclu}), we conclude by summarizing the paper and pitching some ideas for further research.

\section{ EQUATIONS OF MOTION \label{sec2}}

We start by deriving the equations of motion for a rattleback. The following notation will be used. All vectors and matrices are bold-faced. 

\begin{itemize}
\item $M$ for the mass of the rattleback. 
\item $\vect{v}$ for the velocity vector of the centre of mass. 
\item $\vect{f}$ for the force exerted by the table on the body. 
\item $\vect{\omega}$ for the angular velocity of the body. 
\item $\vect{r}$ for the vector from the center of mass to the point of contact. 
\item $\vect{\hat{u}}$ for the vertically upward unit vector. 
\item $\vect{I}$ for Inertia matrix of the rattleback
\end{itemize}

\begin{figure}[htp]
\centering
\includegraphics[width=\linewidth]{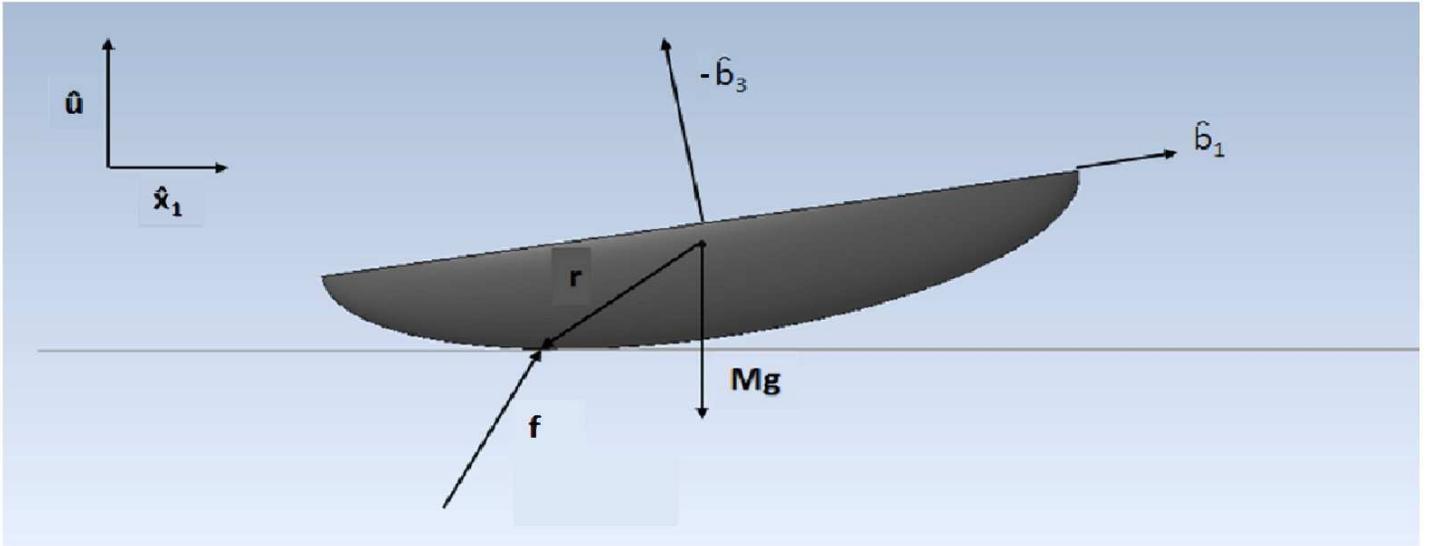} 
\caption{RATTLEBACK WITH FORCES AND AXES }
\label{rattfbd}
\end{figure}

 The set of axes fixed to the rattleback and centered at the center of mass are denoted by $\vect{\hat{b_1}}$, $\vect{\hat{b_2}}$ and $\vect{\hat{b_3}}$ as shown in Fig. \ref{rattfbd}. The inertial system will be fixed in space with $\vect{\hat{u}}$ pointing upwards and $\vect{\hat{x}_1}$ and $\vect{\hat{y}_1}$ in the horizontal plane

The surface acts on the rattleback with a contact force $\vect{f}$ and the relevant equations of motion( for a review of Newtonian mechanics see \cite{goldstein,jose}) can be written as

\begin{align}
\begin{split}
M \frac{\mathrm{d}\vect{v}}{\mathrm{dt}} = \vect{f} - Mg \vect{\hat{u}} \\ \label{newton}
\frac{\mathrm{d}\vect{h}}{ \mathrm{dt}}  = \vect{r} \times \vect{f}
\end{split}
\end{align} 

Note that, throughout this paper, $\frac{\mathrm{d}\vect{p}}{\mathrm{dt}}$ denotes the time derivative of $\vect{p}$ in the inertial frame and time derivative in the moving frame is denoted by $\vect{\dot{p}}$. Since the body fixed frame is rotating with $\vect{\omega}$, the derivatives in the two frames are related as follows

\begin{equation}
\frac{ \mathrm{d}\vect{p}}{\mathrm{dt}}= \vect{\dot{p}} + \vect{\omega} \times \vect{p}
\end{equation}

where $\vect{p}$ is any vector. If the principal inertia's are $I_{11}, I_{22}$ and $I_{33}$ and the offset angle between the geometric and inertia axes is $\delta$, the inertia matrix, in the body fixed frame, can be expressed as 

\begin{equation}
\vect{I} = \begin{bmatrix}  
I_{11} \mathrm{cos}^2 \delta + I_{22} \mathrm{sin}^2 \delta & \frac{(I_{11}-I_{22})}{2} \mathrm{sin}(2 \delta)&0 \\
\frac{(I_{11}-I_{22})}{2} \mathrm{sin}(2 \delta) & I_{11} \mathrm{cos}^2 \delta + I_{22} \mathrm{sin}^2 \delta &0 \\
0& 0 & 1 \end{bmatrix}  \label{matrix}
\end{equation}

The rattleback will be modeled as a semi-ellipsoid and it's shape is given by $q = 0$ where 

\begin{equation}
q = \frac{r_1^2}{a_1^2} + \frac{r_2^2}{a_2^2} + \frac{r_3^2}{a_3^2} - 1
\end{equation}

and $a_1, a_2$ and $a_3$ are the semi principal axes of the rattleback in the body fixed frame. The point of contact, in the body frame,  is $(r_1, r_2, r_3)$ and these form the components of the vector $\vect{r}$. The effects of dissipative forces like friction, air resistance is ignored. The rattleback is assumed to roll without slipping on the surface which implies that the instantaneous point of contact is at rest. Thus, friction is present(in fact, it is necessary) but it does not do any work. 

Since we will be dealing with all quantities in the body fixed frame we can expand the time derivatives as $\frac{\mathrm{d}\vect{p}}{\mathrm{dt}} = \vect{\dot{p}} + \vect{\omega} \times \vect{p}$ and together with Eqn. \ref{newton} to obtain 

\begin{align}
\begin{split}
&\vect{I} \vect{\dot{\omega}} + M\vect{r}\times(\vect{\dot{\omega}} \times \vect{r} ) = M \vect{r} \times(\vect{\dot{r}} \times \vect{\omega}+ \left(\vect{\omega} \times \vect{r}\right) \times \vect{\omega}+ g_c \vect{\hat{u}})  + \vect{I} \vect{\omega} \times \vect{\omega} \\ 
& \vect{\dot{\omega}} = \vect{I}'(u)^{-1} \left( M \vect{r} \times(\vect{\dot{r}} \times \vect{\omega}+ \left(\vect{\omega} \times \vect{r}\right) \times \vect{\omega}+ g_c \vect{\hat{u}})  + \vect{I} \vect{\omega} \times \vect{\omega} \right)  \label{omgdot}
\end{split}
\end{align}

where $\vect{I}'(u) = \vect{I} + M[\vect{r}_{\times}][\vect{r}_{\times}]^T $ and $[\vect{r}_{\times}]$ is the cross product matrix composed of the components of $\vect{r}$. We have written the matrix $\vect{I}'(u)$ as a function of only $u$ because, as will be evident, $\vect{r}$ is only a function of $\vect{\hat{u}}$. Eqn. \ref{omgdot} can be used to integrate angular velocity forward in time.

The equations for propagating unit vector $\vect{\hat{u}}$ can be obtained from \ref{udot}.Since $\vect{\hat{u}},\vect{\hat{x}_1} $ and $\vect{\hat{y}_1}$, are fixed in inertial space, $\frac{\mathrm{d}\vect{u}}{\mathrm{dt}} = 0$, $\frac{{\mathrm{d}\vect{x_1}}}{{\mathrm{dt}}} = 0$ and $\frac{{\mathrm{d}\vect{y_1}}}{\mathrm{dt}} = 0$. And their derivatives in the body fixed frame can be written as 

\begin{align}
\begin{split}
\vect{\dot{u}} &= \vect{\hat{u}} \times \vect{\omega} \\ \label{udot}
\vect{\dot{x_1}} & = \vect{\hat{x_1}} \times \vect{\omega} \\
\vect{\dot{y_1}} &= \vect{\hat{y_1}} \times \vect{\omega} \\
\end{split}
\end{align}

The vectors $\vect{r}$ and $\vect{\dot{r}}$ can be obtained as a function of $\vect{\hat{u}}$ and $\vect{\omega}$ and they depend on the exact shape of the rattleback. The unit vector $\vect{\hat{u}}$, is along the gradient of the surface $q =0$. Thus, $\vect{\hat{u}}$, can be expressed by scaling the gradient. 

\begin{align}
\begin{split}
 \vect{\hat{u}} & = - \frac{\nabla q}{\vert \nabla q \vert} \\
  &=  \frac{-r_1}{a_1^2 \vert \nabla q \vert} \vect{\hat{b_1}} + \frac{-r_2}{a_2^2 \vert \nabla q \vert}\vect{\hat{b_2}} +\frac{-r_3}{a_3^2 \vert \nabla q \vert} \vect{\hat{b_3}} 
 \end{split}  \label{udef}
 \end{align}
 
 where $(r_1, r_2, r_3)$ are the coordinates of the instantaneous point of contact. 
 
 \begin{align}
\begin{split}
 \vert \nabla q \vert = k = \frac{1}{\sqrt{u_1^2a_1^2+u_2^2a_2^2+ u_3^2a_3^2}}
\end{split} \label{kdef}
\end{align}
 
  For the sake of brevity, $\vert \nabla q \vert$, the norm of the gradient, will be henceforth referred to as $k$, as defined in Eqn. \ref{kdef}, not to be confused with $\hat{\vect{k}}$, which is the unit vector in the third inertial direction.  Thus we can, now, find $\vect{r}(u)$ by manipulating Eqn. \ref{udef}. 
 \begin{align}
 \begin{split}
  \vect{r}(u) &=  -k \left( u_1 a_1^2 \vect{\hat{b_1}} + u_2 a_2^2 \vect{\hat{b_2}}  + u_3 a_3^2 \vect{\hat{b_3}} \right)
 \end{split} \label{rdef}
 \end{align}
 
$\vect{\dot{r}(u, \omega)}$ can be similarly obtained by differentiating  $\vect{r(u)}$ to obtain. 
 
 \begin{align}
\begin{split}
\dot{\vect{r}} = \begin{bmatrix}
-a_1^2(k\dot{u_1}+ \dot{k}u_1) \\
-a_2^2(k\dot{u_2}+ \dot{k}u_2) \\
-a_3^2(k\dot{u_3}+ \dot{k}u_3) 
\end{bmatrix}
\end{split} \label{rdotdef}
\end{align}

where $u_1$, $u_2$ and $u_3$ are the components of $\vect{\hat{u}}$. $k$ is defined in Eqn. \ref{kdef} and $\dot{k}$ is the time derivative of $k$ and can be obtained by differentiating Eqn. \ref{kdef} with respect to time. 

Eqn.\ref{omgdot} and Eqn. \ref{udef} will help us integrate $ \dot{\vect{\omega}}$ and $\vect{\dot{u}}$ forward in time for a rattleback rolling on a rough surface under it's own weight and not subjected to any dissipative forces. The formulation of the state equations done here is similar to that in \cite{garcia} and, for further reading, the interested reader is referred to the same.

\section{SPIN RESONANCE \label{sec3}}

Let us simulate the rattleback model developed in section \ref{sec2}.The parameters used for the simulations are listed in Tab. \ref{simulationpara}. All parameters are in SI units. In our simulation,  the rattleback will have a very small initial non-zero pitch angle(and or roll angle) which ensures that the simulated rattleback does not simply rotate for all time about the point of contact exactly below the centre of mass.  Thus, instead of assuming $\vect{\hat{u}}$ initially to be $\vect{\hat{u}} =[0\,\, 0\,\, -1]^T $  we offset it's value ever slightly to  $\vect{\hat{u}} = [.02\,\, .02 \,\,-.9996]^T$. 
  
 \begin{center}
 \captionof{table}{SIMULATION PARAMETERS}
\begin{tabular}{ c | c | c | c }
  \hline                       
  $M$ & $7 \times 10^{-4} \,\, \textrm{kg}$ & $d_{v1}$ & $.65 \times 10^{-1} \,\, \mathrm{N  \,m^{-2} \,s^{2}}$  \\
  $a_1$ & $4.7 \times 10^{-2} \,\, \textrm{m}$  & $d_{v2} $ & $.19\times 10^{-1}\,\, \mathrm{N  \,m^{-2} \,s^{2}}$ \\
 $a_2$ & $1.0 \times 10^{-2} \,\, \textrm{m}$   & $d_{v3} $ & $.175 \times 10^{-1} \,\, \mathrm{N  \,m^{-2} \,s^{2}}$   \\
 $a_3$ &  $ 1.0 \times 10^{-2} \,\, \textrm{m} $  & $d_{\omega1}$ & $.55 \times 10^{-3} \,\, \mathrm{N \, m \, rad^{-2} \, s^{2}}$   \\
 $d_{\omega2} $ & $.03 \times 10^{-3} \,\, \mathrm{N \, m \, rad^{-2} \, s^{2}}$  \\
 $d_{\omega3} $ & $.189 \times 10^{-3} \,\,  \mathrm{N \, m \, rad^{-2} \, s^{2}} $   \\
        \hline  
    \end{tabular} \label{simulationpara}
   \end{center}

   \begin{figure}[htp]
\begin{center}
\begin{tabular}{c}
\subfigure[Plot of Spin rate ($n= \vect{\omega} . \vect{\hat{u}}$)]{\includegraphics[width=\linewidth]{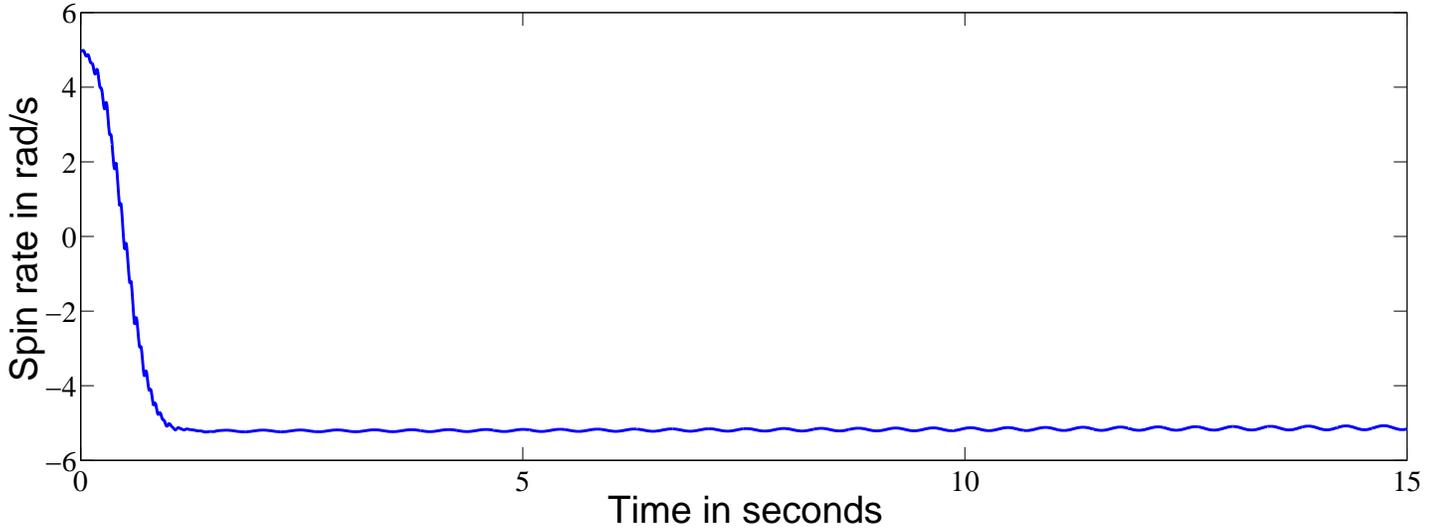} \label{reversal1}} \\
\subfigure[Plot of roll($\omega_1$) and pitch($\omega_2$) angular velocities ]{\includegraphics[width=\linewidth]{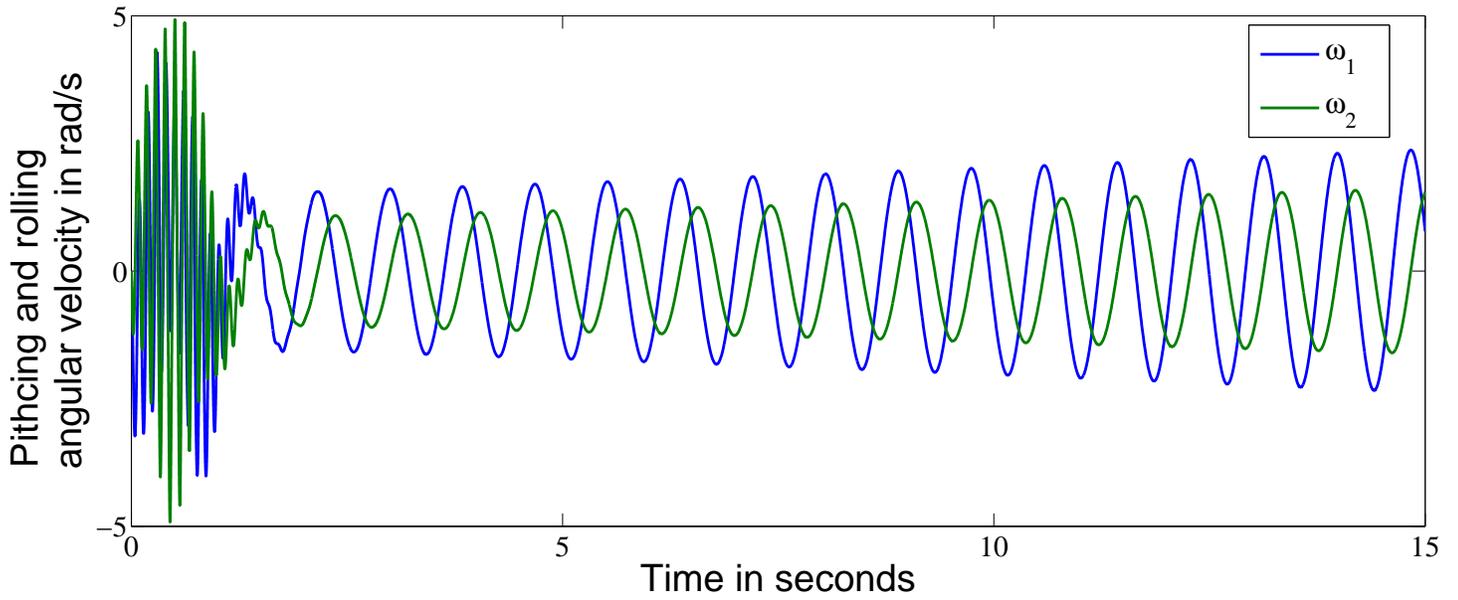}\label{reversal2}}
\end{tabular}
\caption{Plot of (a) spin rate (b) roll and pitch angular velocities with time of a simulated rattleback(conservative model) rotated in the unstable spin direction with initial $\vect{\omega} =[0 \,\,0\,\, -5]^T \mathrm{rad/s}$ and $\vect{\hat{u}} = [.02 \,\,.02\,\, -.9996]^T$.}
\end{center}
\end{figure}

  The rattleback rolls without slipping - the horizontal acceleration of the center of mass is provided by the force of static friction and dissipative forces have been ignored. In Fig. \ref{reversal1} and Fig. \ref{reversal2}, are plotted the spin rate and the pitching and rolling angular velocities respectively. The simulated rattleback is spun in the unstable direction. Note the spin reversal which causes the rattleback to reverse from an initial spin of $5 \,\, \mathrm{rad/s}$ to $-5 \,\, \mathrm{rad/s}$. The pitching and rolling angular velocities, it appears, have a high frequency component that appears only during the reversal and dies down subsequently leaving only monochromatic oscillations. For detailed numerical simulations of spin reversal using conservative and dissipative models of the rattleback, the interested reader is referred to \cite{garcia}. 
   
In this paper, we are specifically interested in the dynamic behavior of the rattleback when the platform undergoes simple harmonic oscillations. The apparent acceleration due to gravity , $g_{c,app}$, in the body-fixed frame ( $\vect{\hat{b}_1},\vect{\hat{b}_2}$ and $\vect{\hat{b}_3}$ ) can be written as 
\begin{align}
g_{c,app} = g_c + A_0\omega_0^2 \mathrm{cos}(\omega_0 t) \\
\end{align}

\begin{figure}[htp]
\centering
\includegraphics[width=\linewidth]{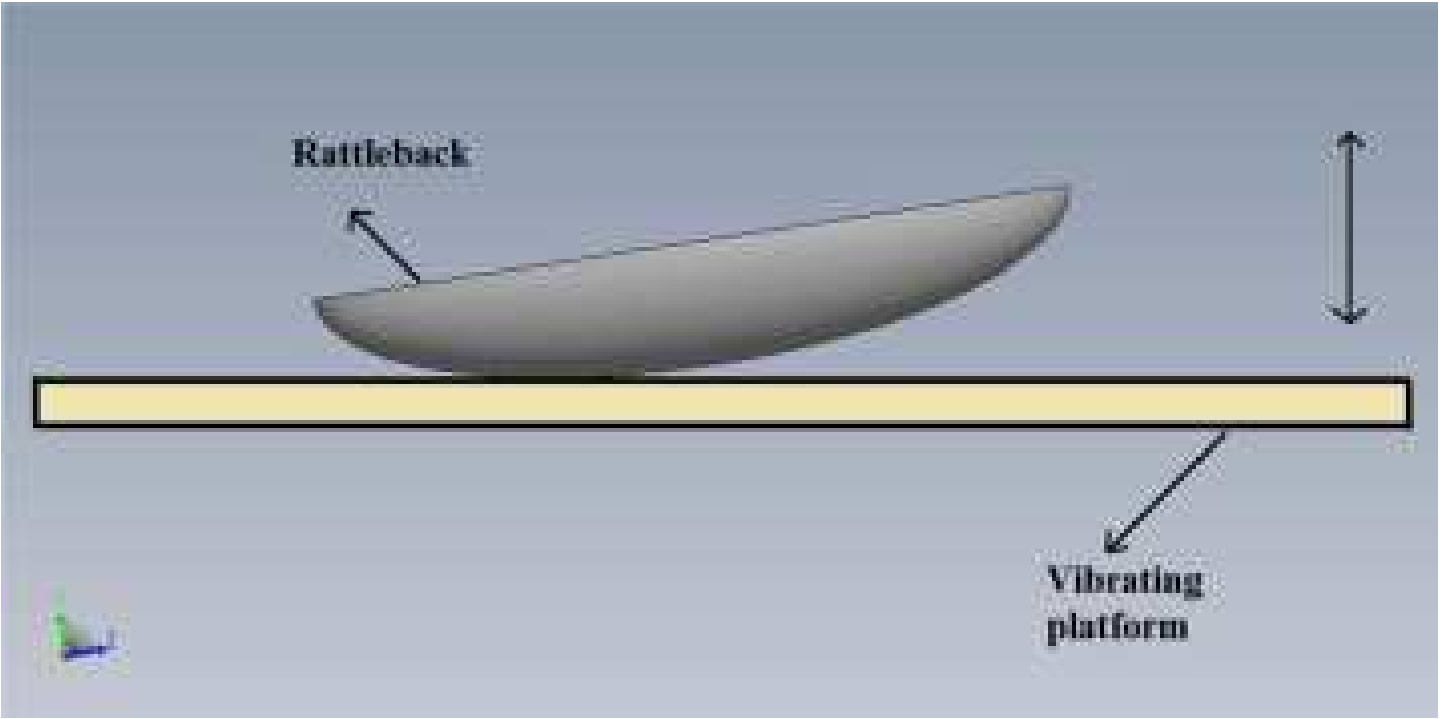}
\caption{ A rattleback rolling on a harmonically oscillating platform }
\label{vib}
\end{figure}

$g_{c,app}$ is the apparent acceleration due to gravity, $A_0$ and $\omega_0$ are the amplitude and angular velocity corresponding to the platform vibration respectively.  Also, $\omega_0 = 2 \pi f_{base}$. To investigate the effect of damping, simulations were run using the conservative and the dissipative model . The amplitude of platform oscillations was varied with frequency such that the magnitude of change in acceleration in the frame of the rattleback was $ 3 \,\, \mathrm{m\, s^{-2}}$ at all frequencies. Please see Fig. \ref{vib}.  

\begin{figure}[htp]
\centering
\includegraphics[width=\linewidth]{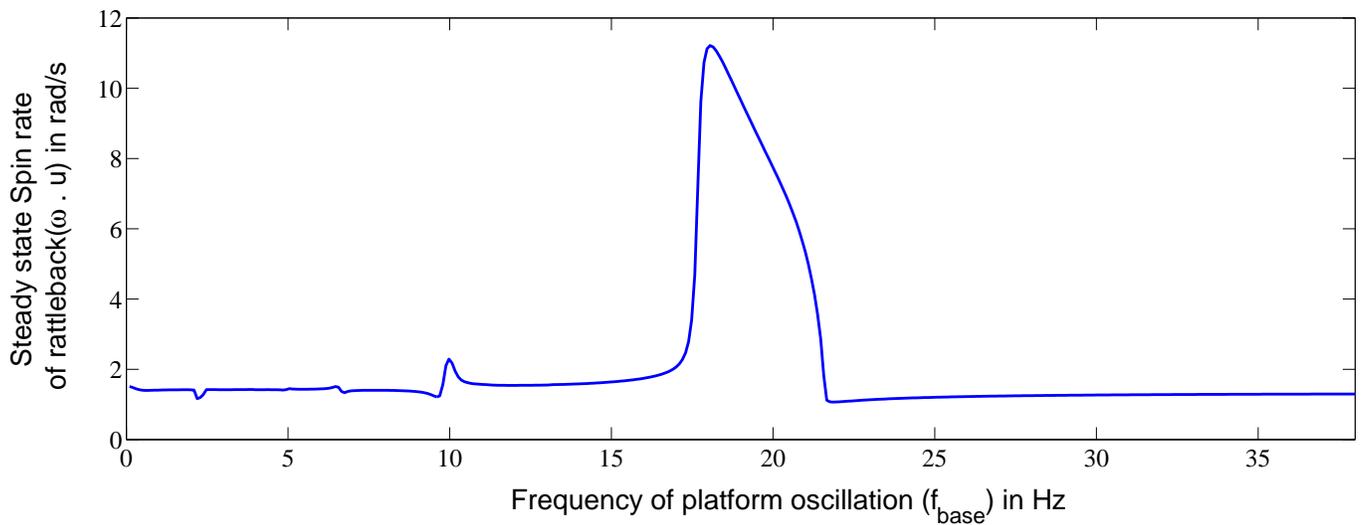}
\caption{ Plot of steady state spin rate($n$) of a simulated rattleback(Conservative model) against frequency of platform oscillation with initial conditions : $\vect{\hat{u}}= \left[.02 \,.02\, .9996 \right]^T$  and  $\vect{\omega} = \left[0 \,0\, 0 \right]^T $ }
\label{spinresonance1}
\end{figure}

 Simulations were run with the frequency($f_{base}$) of platform oscillations varying from $0 $ to $38 \,\, \mathrm{Hz}$ and the conservative model was used( drag forces were ignored).   The variation in steady state spin rate  ($n = - \vect{\omega} . \vect{\hat{u}}$, as recorded at $t = 10 \,\, \mathrm{s}$) with the frequency of platform oscillations is plotted in Fig. \ref{spinresonance1} .  There is a clear resonance at $f_{base}  \approx 19 \,\, \mathrm{Hz}$. The rattleback achieves a high spin angular velocity of $11.5 \,\, \mathrm{rad/s}$. 
 
\begin{figure}[htp]
\centering
\includegraphics[width=\linewidth]{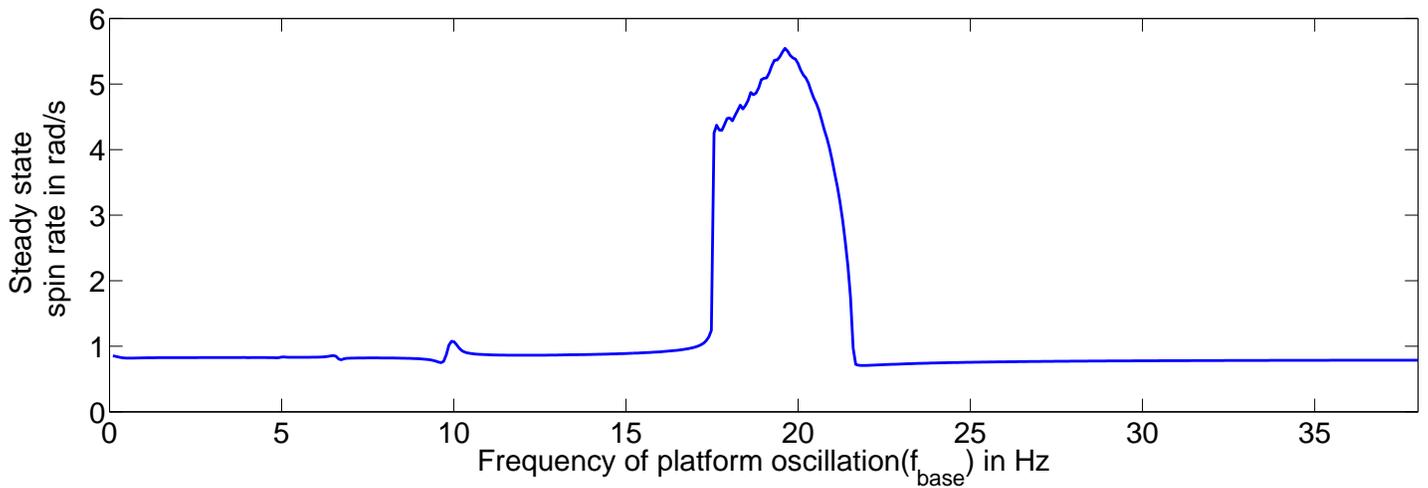}
\caption{ Plot of steady state spin rate($n$) of a simulated rattleback(Dissipative model) against frequency of platform oscillation with initial conditions : $\vect{\hat{u}}= \left[.02 \,.02\, .9996 \right]^T $  and $\vect{\omega} = \left[0 \,0\, 0 \right]^T $ }
\label{spinresonance2}
\end{figure}

We now seek to include dissipation in the analysis.  To include dissipation, aerodynamic drag forces and torques were modeled as being proportional to the square of velocity and angular velocity respectively. Mathematically, 

\begin{align*}
\vect{f_d} &= - \begin{bmatrix} d_{v_1}& 0 & 0\\ 0 & d_{v_2} & 0 \\ 0& 0& d_{v_3} \end{bmatrix} \begin{bmatrix} \vert v_1 \vert v_1 \\ \vert v_2 \vert v_2 \\ \vert v_3 \vert v_3 \end{bmatrix} \\
\vect{\tau_d} &=  - \begin{bmatrix} d_{\omega_1}& 0 & 0\\ 0 & d_{\omega_2} & 0 \\ 0& 0& d_{\omega_3} \end{bmatrix} \begin{bmatrix} \vert \omega_1 \vert \omega_1 \\ \vert \omega_2 \vert \omega_2 \\ \vert \omega_3 \vert \omega_3 \end{bmatrix} 
\end{align*}

The values of the parameters $d_{v_1}$, $d_{v_2}$ etc. is shown in Tab. \ref{simulationpara}.  The results were similar to that for the conservative model as can be observed in Fig. \ref{spinresonance2}. The resonance peak is smaller which is not surprising as energy is siphoned off from the rattleback by the drag forces. The shape of the peak has changed slightly and the resonant frequency has shifted slightly to the left. 

\section{ THEORETICAL ANALYSIS\label{sec4}}
 
So far we have seen that the rattleback has a spin resonance with respect to base vibrations. If the base on which the rattleback is placed starts undergoing vibrations then in the neighborhood of a certain frequency it will start spinning. We do not have any insight or understanding of the mechanism that causes this resonance. Before we go further, we need to take a closer look at the response of the rattleback. 
 
 \begin{figure}[htp] 
\centering
\includegraphics[width=\linewidth]{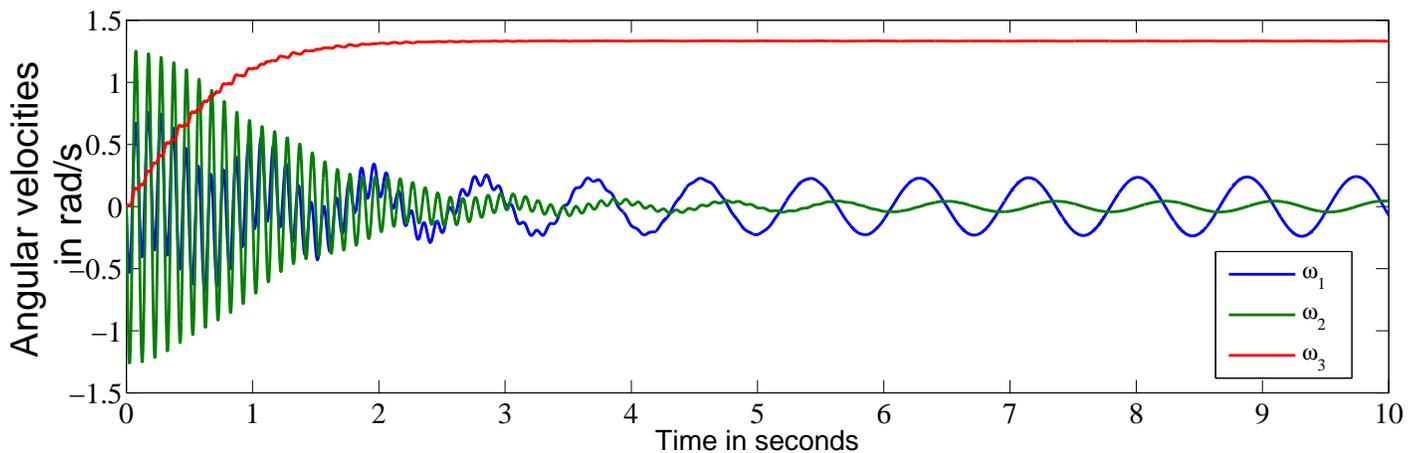}  
\caption{ Plot of Angular velocities against time of a simulated rattleback with initial  $\vect{\omega} =[0 \,\,0\,\, 0]^T rad/s$ and $\vect{\hat{u}} = [.02 \,\,.02\,\, -.9996]^T$.  }
\label{omega3}
\end{figure} 

  See Fig. \ref{omega3} which plots the angular velocities of a simulated rattleback with zero initial spin ($\vect{\omega} = [ 0 \ 0 \ 0]^T$) and initial orientation slightly offset from the equilibrium position ($\vect{\hat{u}} = [.02 \,\,.02\,\, -.9996]^T$) with respect to time. The rolling and pitching angular velocities oscillate about zero and the spin angular velocity increases and reaches a steady state value as expected. It is clear that the the pitching and rolling angular velocities have two frequency components - a high frequency component that diminishes as soon as the spin rate ceases increasing and a slower frequency that does not seem to undergo any attenuation. Our proposition here is that \emph{only the high frequency is coupled with the spinning motion of the rattleback}. The lower frequency does not have any coupling(very little coupling, to be precise ) with spinning. 
 
  Consider a stationary rattleback. If we impart some pitching motion to the rattleback, both the aforementioned frequencies will appear in the pitching and rolling responses as shown in Fig. \ref{omega3} .  Simultaneously, the spin of the rattleback will increase. However, only the fast frequency will undergo decay. As spin increases, more and more energy from the fast frequency is siphoned off into the spinning and the amplitude corresponding to the same decreases and ultimately becomes zero(or negligibly small). The slow frequency is not coupled with the spinning and therefore it does not die down as spin increases. Note that the decay is not due to any damping or dissipation as none has been considered in this particular model. Instead, it is because of the flow of energy from the fast frequency into the spinning motion. Bondi in \cite{bondi} commented on this behaviour by analyzing the motion of roots of the characteristic equation in the complex plane but \textit{the crucial insight here is that of the two frequencies appearing in the response only one is coupled to the spinning. }
  
  Henceforth, the two frequencies will be referred to as the \textit{Coupled frequency }(the faster frequency for the rattleback described above) and the \textit{Uncoupled frequency}. Reversals can be similarly explained. Consider a rattleback spun in the unstable direction. Within a short time, pitching oscillations build up and the spin rate starts decreasing. The amplitude of the coupled frequency reaches a peak and then starts decreasing concurrent with the spin rate dropping to zero and reversing to spin in the stable direction.  

The coupling of the coupled frequency with the spinning motion is an important insight because we found that \textit{Spin resonance occurs when the platform oscillations have a frequency that is equal to (or in the neighborhood of)twice the coupled frequency.} Mathematically, if we represent by $f_{coupled}$ the coupled frequency then 

\begin{align}
\begin{split}
f_{base,resonant} = 2 f_{coupled,} 
\end{split} \label{resonant}
\end{align}

\begin{figure}[htp] 
\includegraphics[width=.9\linewidth]{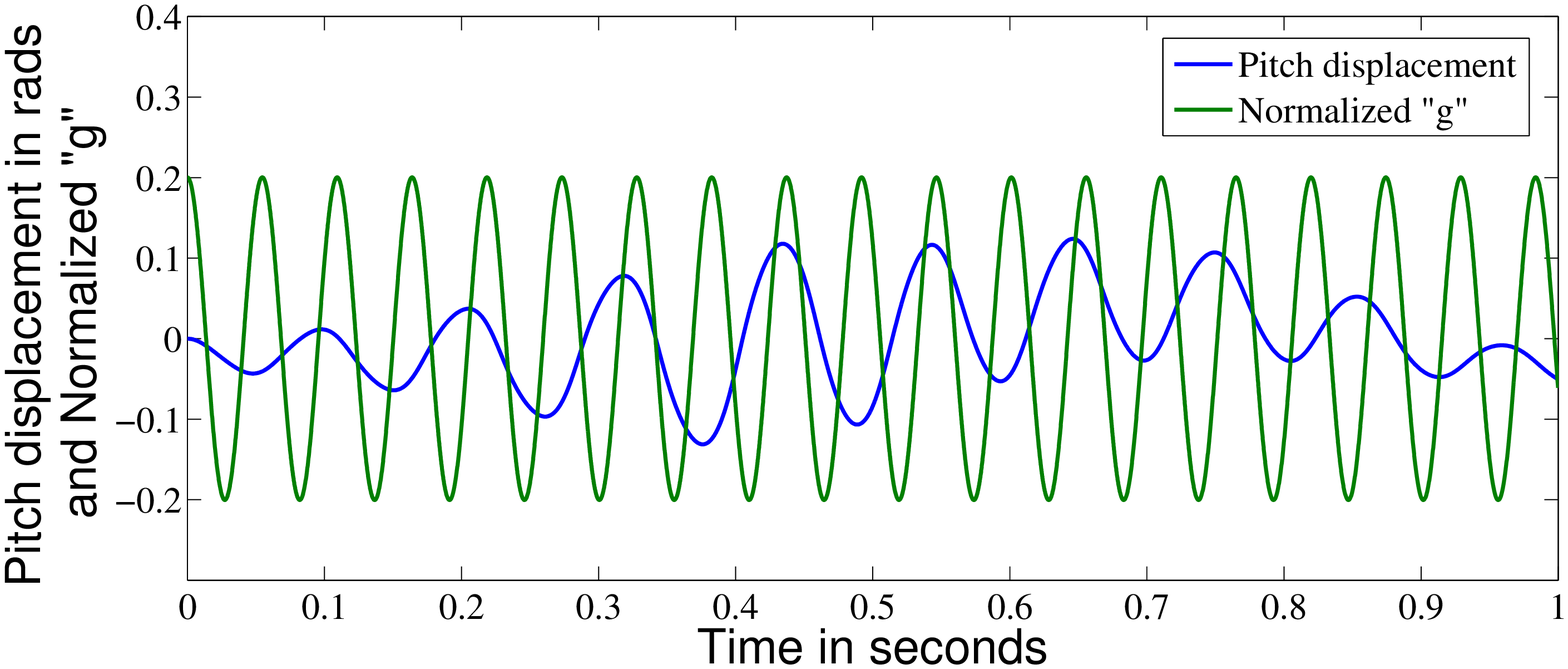}  
\caption{ Plot of the scaled acceleration due to gravity and pitch displacement against time with initial  $\vect{\omega} =[0 \,\,0\,\, 0]^T \mathrm{rad/s}$ and $\vect{\hat{u}} = [.02 \,\,.02\,\, -.9996]^T$ and $f_{base} = \, 19 \,\,\mathrm{Hz}$ (Spin resonant frequency for the rattleback).  }
\label{pitchAndg}
\end{figure}

This equation makes a lot of intuitive sense because when Eqn. \ref{resonant} holds the pitch motion and the base oscillation become phase locked with each other such that when the apparent acceleration due to gravity reaches it's maximum value the pitch displacement is maximum. Since \textit{in one time-period of the coupled frequency, the magnitude of pitch displacement will reach maxima twice(once to each side), for resonance it is necessary that $g_{c,app}$ attain maxima at the same times and this is possible when base oscillations are twice as fast}. This can be observed in Fig. \ref{pitchAndg} which plots the pitch displacement and the scaled apparent acceleration due to gravity with respect to time for a simulated rattleback kept on a platform vibrating at $f_{base}= 19 \,\, \mathrm{Hz}$ ( the resonant frequency for the rattleback with parameters in Tab. \ref{simulationpara}). 

Since the described frequencies appear when the rattleback is given small perturbations in pitch(and roll) from the zero equilibrium positions, we reasoned that it might be possible to gain some understanding of the same by linearization around the equilibrium point. As it turns out, in the linear model of the rattleback, the two frequencies appear as Eigenvalues of the Jacobian matrix. In this section, we will obtain a linearized model around the equilibrium point and obtain expressions for the coupled and uncoupled frequencies. Later, this will be validated by running simulations of rattlebacks with different parameters. 

\subsection{Linearization \label{linearization}}

Consider a general system with $\vect{x} \in \vect{\mathbb{R}^{n \times 1}}$ as the state vector. 
 
\begin{align}
\begin{split}
\frac{\mathrm{d}\vect{x}}{\mathrm{dt}} = \vect{F(x, t)}
\end{split}
\end{align}
 
  This system will have a equilibrium point $\vect{x_0}$ if $\vect{F(x_0)}=0$ and let us assume, without loss of generality, that $\vect{x_0}= 0$.  The derivative, $\vect{J}=\frac{\vect{\partial F}}{\vect{\partial x}}$ is called the Jacobian and it is a $n \times n$ real matrix. Consider the governing equations of the rattleback derived in section \ref{sec2}. Our state vector, $\vect{x}$ will be the augmentation of $\vect{\omega}$ and $\vect{\hat{u}}$ and 
  

\begin{align}
\begin{split}
&\vect{F}= \begin{bmatrix}
\vect{F_{\omega}} \\ \vect{F_{u}}
\end{bmatrix} \\
&=
\begin{bmatrix}
 \vect{I}'(u)^{-1} \bigl[  M \vect{r} \times(\vect{\dot{r}} \times \vect{\omega}+ \left(\vect{\omega} \times \vect{r}\right) \times \vect{\omega}+ g_c \vect{\hat{u}}) 
 + I \vect{\omega} \times \vect{\omega} \bigr]\\
 \vect{\hat{u}} \times \vect{\omega}
\end{bmatrix}
\end{split}
\end{align}

In this formulation, we are interested in linearizing the system about the equilibrium point $\vect{x_0} = [0\,\,0\,\, 0\,\, 0\,\, 0\,\, -1]^T$- i.e, all the angular velocities are zero( $\vect{\omega} = [0 \ 0 \ 0]^T$) and the orientation is vertically upwards ($\vect{\hat{u}} = [ 0 \ 0 \ -1]^T$ ). The Jacobian $\vect{J}$ can, then, be written as 

\begin{align}
\vect{J} = \begin{bmatrix}
\frac{\vect{\partial} \vect{F_\omega}}{\vect{\partial} \vect{\omega}} & \frac{\vect{\partial}\vect{ F_\omega}}{\vect{\partial} \vect{\hat{u}}} \\
\frac{\vect{\partial}\vect{ F_u}}{\vect{\partial} \vect{\omega}} & \frac{\vect{\partial}\vect{ F_u}}{\vect{\partial} \vect{\hat{u}}} 
\end{bmatrix}
\end{align}

These derivatives will be evaluated one after another as follows.  For $\frac{\vect{\partial} \vect{F_\omega}}{\vect{\partial} \vect{\omega}}$ we can write

\begin{align}
\frac{\vect{\partial} \vect{F_\omega}}{\vect{\partial} \vect{\omega}}& =\vect{I}'(u)\frac{\vect{\partial} }{\vect{\partial} \vect{\omega}} \left( M \vect{r} \times(\vect{\dot{r}} \times \vect{\omega}+ \left(\vect{\omega} \times \vect{r}\right) \times \vect{\omega}+ g_c \vect{\hat{u}})  + \vect{I} \vect{\omega} \times \vect{\omega} \right) 
\end{align}

 All the components of the terms $M\vect{r} \times \left(\vect{\omega} \times \vect{r}\right) \times \vect{\omega}$ and $\vect{I} \vect{\omega} \times \vect{\omega}$ will contain only 2nd order terms of $\omega$ and will therefore vanish.   The term $M \vect{r} \times \vect{\dot{r}} \times \vect{\omega}$ can also be shown to be second order in $\vect{\omega}$. Thus, we can write 
\begin{align}
\frac{\vect{\partial} \vect{F_\omega}}{\vect{\partial} \vect{\omega}} = \begin{bmatrix}
0&0&0 \\ 
0&0&0 \\
0& 0&0 
\end{bmatrix}
 \label{df1domg}
\end{align}
Let us consider, now, $\frac{\vect{\partial} \vect{F_\omega}}{\vect{\partial} \vect{\hat{u}}}$. We can write 

\begin{align}
\begin{split}
\frac{\vect{\partial} \vect{F_\omega}}{\vect{\partial} \vect{\hat{u}}} = \vect{I}'(u)\frac{\vect{\partial} }{\vect{\partial} \vect{\hat{u}}} \left( M \vect{r} \times(\vect{\dot{r}} \times \vect{\omega}+ \left(\vect{\omega} \times \vect{r}\right) \times \vect{\omega}+ g_c \vect{\hat{u}})  + \vect{I} \vect{\omega} \times \vect{\omega} \right) +\\
\frac{ \vect{\partial}}{\vect{\partial} \vect{\hat{u}}} \left(\vect{I}'(u)\right) \left( M \vect{r} \times(\vect{\dot{r}} \times \vect{\omega}+ \left(\vect{\omega} \times \vect{r}\right) \times \vect{\omega}+ g_c \vect{\hat{u}})  + \vect{I} \vect{\omega} \times \vect{\omega} \right) 
\end{split}  \label{df1du}
\end{align}

The second term in Eqn. \ref{df1du} can be dropped because we are evaluating this point at the equilibrium point and $F_{\omega}(0)= 0$.  The only non-zero contribution comes from $Mg_c \vect{r} \times \vect{\hat{u}}$ and after some algebra, we can write 

\begin{align}
\begin{split}
&\frac{ \vect{\partial}}{\vect{\partial} \vect{\hat{u}}} Mg_c (\vect{r} \times \vect{\hat{u}}) \\ 
&= Mg_c \begin{bmatrix}
0 & -u_3(a_2^2 - a_3^2)k & 0\\
-u_3(a_3^2- a_1^2) k & 0& 0\\
0 & 0 & 0 
\end{bmatrix} \label{df1du2}
\end{split}
\end{align}

If we evaluate $\vect{I}'(u)$ at $u= [0 \,\,0\,\, -1]^T$ to get $\vect{I}(\vect{u_0})$ and combining equations Eqn. \ref{df1du} and Eqn. \ref{df1du2} we obtain 

\begin{align}
\begin{split}
\frac{ \vect{\partial} \vect{F_{\omega}}}{\vect{\partial} \vect{\hat{u}}}  = \vect{I}'(u_0)^{-1} \begin{bmatrix}
0 & -u_3(a_2^2 - a_3^2)k & 0\\
-u_3(a_3^2- a_1^2) k & 0& 0\\
0 & 0 & 0 
\end{bmatrix}
\end{split}
\end{align}

For brevity and to avoid very long expressions let us denote the diagonal elements of the inertia matrix I by $I_1$, $I_2$, $I_3$ and the nonzero off-diagonal elements by $I_{12}$. Note that the inertia matrix I was defined in Eq. \ref{matrix} in terms of the principal moments of inertia $I_{11}$, $I_{22}$ and $I_{33}$ and is reproduced here for convenience of the reader as 

\begin{align}
\begin{split}
\vect{I} = & \begin{bmatrix}
I_{11} \mathrm{cos}^2 \delta + I_{22} \mathrm{sin}^2 \delta & \frac{(I_{11}-I_{22})}{2} \mathrm{sin}(2 \delta)&0 \\
\frac{(I_{11}-I_{22})}{2} \mathrm{sin}(2 \delta) & I_{11} \mathrm{cos}^2 \delta + I_{22} \mathrm{sin}^2 \delta &0 \\
0& 0 & 1 \end{bmatrix} \\ 
&= \begin{bmatrix}
I_1& I_{12} &0  \\
I_{12}& I_2 & 0 \\
 0& 0 & I_3\\
\end{bmatrix}
\end{split}
\end{align}

  Remembering that $\vect{I}'(u)= \vect{I} + M [\vect{r_{\times}}] [\vect{r_{\times}}]^T$ and substituting for $u_3= -1$, $r_3= a_3$ and $k= 1/a_3$ we can write the final form of Eqn. \ref{df1du} as

\begin{align}
\begin{split}
&\frac{ \vect{\partial} \vect{F_{\omega}}}{\vect{\partial} \vect{\hat{u}}} \\ 
&=  \frac{Mg_c} {\mathrm{det}(\vect{I}(u_0))} \begin{bmatrix}
-I_{12} I_3 \frac{(a_3^2- a_1^2)}{a_3} & (I_2+Ma_3^2)I_3\frac{(a_2^2- a_3^2)}{a_3} & 0 \\
I_3(I_1+ Ma_3^2 )\frac{a_3^2-a_1^2}{a_3}  & -I_{12} I_3 \frac{(a_2^2- a_3^2)}{a_3} & 0\\
0 & 0 & 0
\end{bmatrix}  \label{df1dufull}
\end{split}
\end{align} 

where $\mathrm{det(.)} $ denotes the determinant. Let us, now, consider the submatrices corresponding to $\vect{F_u}$.  We know that $\vect{F_u} = \vect{\hat{u}} \times \vect{\omega}$.  Taking the derivatives becomes straightforward and they can be written as 

\begin{align}
\begin{split}
\frac{\vect{\partial} \vect{F_u}}{\vect{\partial} \vect{\omega}} & = [\vect{u_\times}] = \begin{bmatrix} 0 & -u_3 & u_2\\ u_3 & 0&  -u_1 \\-u_2 & u_1 & 0 \end{bmatrix} \\
\frac{\vect{\partial} \vect{F_u}}{\vect{\partial} \vect{\hat{u}}} & = [\vect{\omega_\times}]= \begin{bmatrix} 0 & -\omega_3 & \omega_2\\ \omega_3 & 0 & -\omega_1 \\-\omega_2 & \omega_1 & 0 \end{bmatrix} \\
\end{split} \label{df2domg}
\end{align} 

Now we are in a position to write down the Jacobian matrix by combining the expressions for all the submatrices derived in Eqn. \ref{df1domg}, Eqn. \ref{df1dufull} and Eqn. \ref{df2domg}. 

\begin{align}
\begin{split}
\vect{J} = \begin{bmatrix}
0 & 0 & 0 & J_{u_1}^{(1)} & J_{u_2}^{(1)} & 0 \\
0 & 0 & 0 & J_{u_1}^{(2)} & J_{u_2}^{(2)} & 0 \\
0 & 0 & 0 &0 & 0 & 0 \\  \label{jacobian}
0 & 1 & 0 &0 & 0 & 0 \\
-1 & 0 & 0 &0 & 0 & 0 \\
0 & 0 & 0 &0 & 0 & 0 
\end{bmatrix}
\end{split}
\end{align}

where the terms $J_{u_1}^{(1)}$ , $J_{u_1}^{(1)}$ etc are the non-zero terms of the submatrix $\frac{\vect{\partial} \vect{F_\omega}}{\vect{\partial} \vect{\hat{u}}}$ appearing in Eqn. \ref{df1dufull} and, for the sake of completeness, are reproduced here

\begin{align}
\begin{split}
J_{u_1}^{(1)} &= \frac{Mg_c \left( -I_{12} I_3 \frac{(a_3^2- a_1^2)}{a_3} \right)} {I_3(I_1+Ma_3^2)(I_2+ Ma_3^2)- I_{12}^2} \\ \\
J_{u_1}^{(2)} &= \frac{Mg_c \left(I_3(I_1+ Ma_3^2 )\frac{a_3^2-a_1^2}{a_3} \right)} {I_3(I_1+Ma_3^2)(I_2+ Ma_3^2)- I_{12}^2}  \\ \\
J_{u_2}^{(1)} &= \frac{Mg_c \left( (I_2+Ma_3^2)I_3\frac{(a_2^2- a_3^2)}{a_3} \right)} {I_3(I_1+Ma_3^2)(I_2+ Ma_3^2)- I_{12}^2}  \\ \\ 
J_{u_2}^{(2)} &= \frac{Mg_c  \left(-I_{12} I_3 \frac{(a_2^2- a_3^2)}{a_3}  \right)} {I_3(I_1+Ma_3^2)(I_2+ Ma_3^2)- I_{12}^2} 
\end{split} \label{Ju12}
\end{align}

So far we have linearized the governing equations of the rattleback about the equilibrium point $\vect{x_0} = [0\,\,0\,\, 0\,\, 0\,\, 0\,\, -1]^T$ and obtained the Jacobian. We saw that when the rattleback was disturbed from this equilibrium position two frequencies dominated the response. 

To find the eigenvalues of the Jacobian we solve $\mathrm{det}(\vect{J}-\lambda \vect{I})= 0 $; $\lambda$ being the eigenvalues and $\vect{I} $ , not to be confused with the inertia matrix, is the identity matrix .  The characteristic equation can be obtained as 

\begin{align}
\begin{split}
\lambda^2 \left( \lambda^4- \lambda^2 (J_{u_1}^{(2)} -J_{u_2}^{(1)} ) + (J_{u_1}^{(1)}J_{u_2}^{(2)}- J_{u_2}^{(1)}J_{u_1}^{(2)} ) \right) = 0 
\end{split} \label{characteristic}
\end{align}

Solving the characteristic equation will give us the Eigenvalues. Two of the Eigenvalues are zero which is expected since we have two zero columns in the Jacobian matrix corresponding to dynamics of $\omega_3$ and $u_3$. 

\begin{align}
\begin{split}
\lambda_{1,6 } = 0
\end{split}
\end{align}

This implies that the coupling of the spin motion with the coupled frequency is a higher order dynamic phenomenon and will not appear in the linearized model.  This leaves us with a biquadratic equation which can be solved and, for the remaining four eigenvalues, after some manipulation, we obtain  

\begin{align}
\begin{split}
 \lambda_{2,3} &= \pm \sqrt{\frac{J_{u_1}^{(2)} -J_{u_2}^{(1)} + \sqrt{(J_{u_1}^{(2)} +J_{u_2}^{(1)})^2 - 4J_{u_1}^{(1)}J_{u_2}^{(2)} }} {2}} \\ \\
 \lambda_{4,5} &= \pm \sqrt{\frac{J_{u_1}^{(2)} -J_{u_2}^{(1)} - \sqrt{(J_{u_1}^{(2)} +J_{u_2}^{(1)})^2 - 4J_{u_1}^{(1)}J_{u_2}^{(2)} }} {2}}
\end{split}  \label{eigenequation}
\end{align}

 The four values of $\lambda$ will be purely imaginary if evaluated for a real rattleback and this completes our derivation.  If we evaluate Eqn. \ref{eigenequation} for the rattleback considered previously(parameters in Tab. \ref{simulationpara}) then we obtain the eigenvalues below. 
 
 \begin{align}
\begin{split}
\lambda_{1} &= 0 \\
\lambda_{2,3} &= \pm \,\mathrm{j} 63.225 \\
\lambda_{4,5}&= \pm \, \mathrm{j} 7.24 \\
\lambda_{6}&= 0 \\
\end{split} \label{ratt1_0}
\end{align} 

We also obtain six eigenvectors which are complex conjugates of each other. Note that in the original non-linear system, the rattleback's angular velocity around the third axis $\omega_3$, is highly coupled with oscillations in pitch.  This implies that although the coupled and uncoupled frequencies appear in the linearized model of the rattleback, the coupling with $\omega_3$ does not.  Indeed, all the interesting dynamics of $u_3$ and $\omega_3$ are non-linear and we need higher order derivatives of the state equation to describe them accurately.

\subsection{ Validation of Analysis \label{validate}}

In this section, we will demonstrate the efficacy of the linear model in accurately predicting the coupled and uncoupled frequencies of the  rattleback. We will apply the derived formulae and compare the frequencies by running simulations of two rattlebacks with different mass and inertia parameters. 

\subsubsection{Rattleback - I}
Let us consider the rattleback with parameters shown in Tab. \ref{simulationpara} and we evaluate Eqn. \ref{eigenequation} to get the following eigenvalues

\begin{align}
\begin{split}
\lambda_{1,2} &= \pm \,\mathrm{j} 63.225 \\
&\implies f_1 = \frac{\mathrm{abs}(\lambda_{1,2}) }{2 \pi} = 10.06 \,\, \mathrm{Hz}\\
\lambda_{3,4} &= \pm \, \mathrm{j} 7.24 \\
&\implies f_2 = \frac{\mathrm{abs}(\lambda_{1,2}) }{2 \pi} = 1.15 \,\, \mathrm{Hz}\\
\end{split} \label{ratt1eig}
\end{align}

where $\mathrm{j}= \sqrt{-1}$ and $\mathrm{abs}(.)$ represents the Absolute value function. A discrete fourier transform of the pitching angular velocity of the simulated rattleback reveals that the Linearized model derived in subsection \ref{linearization} is accurate at predicting the two fundamental frequencies that appear in the response. 

 \begin{figure}[htp]
\centering
\includegraphics[width=\linewidth]{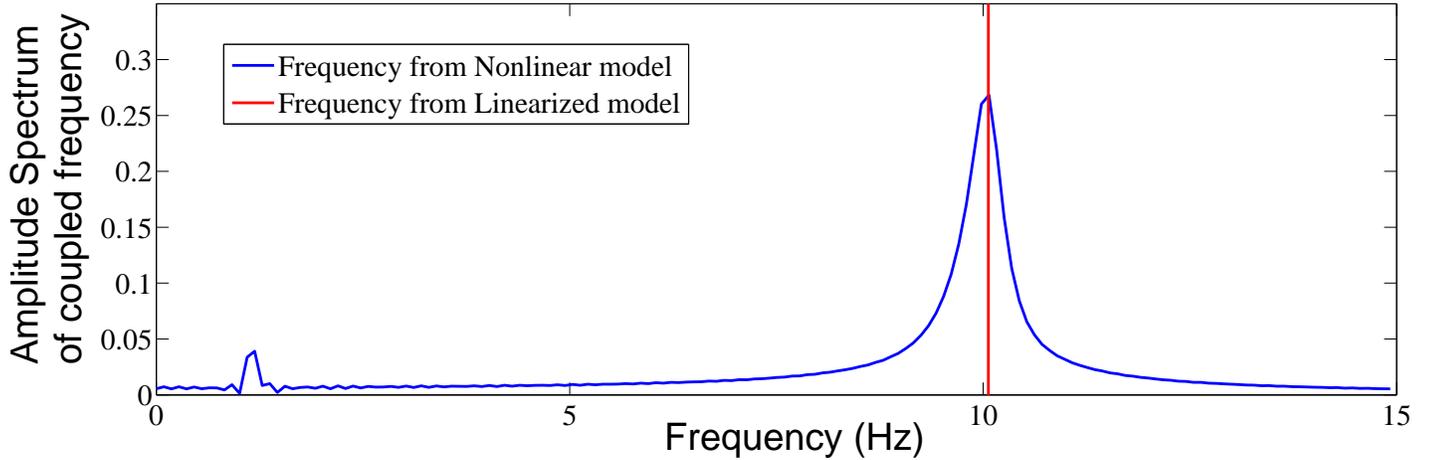}
\caption{ Amplitude spectrum of coupled frequency of a simulated rattleback with initial  $\vect{\omega} =[0 \,\,0\,\, 0]^T \mathrm{rad/s}$ and $\vect{\hat{u}} = [.02 \,\,.02\,\, -.9996]^T$.  }
\label{ratt1compare1}
\end{figure} 

\begin{figure}[htp]
\centering
\includegraphics[width=\linewidth]{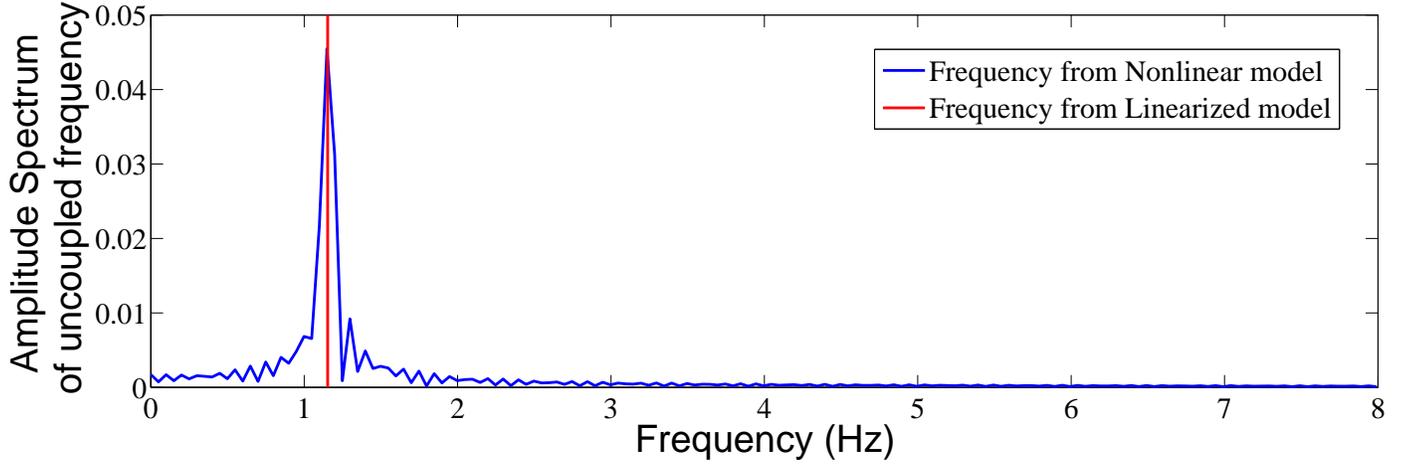}
\caption{ Amplitude spectrum of uncoupled frequency of a simulated rattleback with initial $\vect{\omega} =[0 \,\,0\,\, 0]^T \mathrm{rad/s}$ and $\vect{\hat{u}} = [.02 \,\,.02\,\, -.9996]^T$.  }
\label{ratt1compare2}
\end{figure}

Shown in Fig.\ref{ratt1compare1} is the Amplitude spectrum of the angular velocity of the simulated rattleback. The red line is the predicted frequency value for the coupled frequency obtained from Eqn.\ref{eigenequation} . In Fig.\ref{ratt1compare2} is plotted the amplitude spectrum of the uncoupled frequency and again it can be seen that the value of the frequency predicted by the linearized model is almost exactly equal to that obtained from the full order model. 

Also as can be seen in Fig. \ref{ratt1compare1}, the coupled frequency for this rattleback $10.06 \mathrm{Hz}$ . Using Eqn. \ref{resonant} then, we obtain a value for the resonant frequency as $20.12 \, \mathrm{Hz}$ and this is clearly the case from the results presented in Fig. \ref{spinresonance1} and Fig. \ref{spinresonance2} . 

 Let us linearize the equations about $\vect{x_0}= [0 \, 0 \,.5\, 0\, 0\, -1]$(the rattleback has non-zero $\omega_3$ angular velocity in the stable direction and the orientation is vertically upwards) . Note that Eqn. \ref{jacobian} will not be applicable here so we will use numerical methods. For linearization about stable spin, we obtain the following eigenvalues: - 

\begin{align}
\begin{split}
 \lambda_{1} &= 0 \\
\lambda_{2,3} &= -.4558 \,\pm \,\mathrm{j}63.226 \\
\lambda_{4,5}&= .006 \,\pm \, \mathrm{j}7.246 \\
\lambda_6&= 0 \\
\end{split} \label{ratt1_.5}
\end{align}

Note that the eigenvalues which were previously purely imaginary when evaluated about zero spin(see Eqn. \ref{ratt1_0}) now have real parts. Especially interesting is the fact that the real parts corresponding to the coupled frequency are negative in sign which makes intuitive sense as the rattleback has stable spin. The uncoupled frequency has a small positive real part and this \textit{will cause the amplitude to increase slowly}- this behaviour is a property of Zone I rattlebacks and was observed and commented on by Garcia \textit{et.al}  in \cite{garcia} and can be observed in the simulations we have presented (See Fig. \ref{reversal1} (b) ).  Also note that the imparted angular velocity($\omega_3 =.5 \,\, \mathrm{rad/s}$) does not significantly change the imaginary part of the eigenvalues(compare with the values in Eqn. \ref{ratt1_0} ) which corresponds to the frequencies of oscillation.  The fact that the eigenvalues corresponding to coupled frequency have large negative real parts, unlike the uncoupled frequency, can be used to distinguish the two when linearization is carried out about zero spin in which case the real parts are, of course, zero. 

Linearization about unstable $\omega_3$ ( $\vect{x_0}= [0\, 0\, -.5\, 0\, 0\, -1]$ ) yields the following eigenvalues. 

\begin{align}
\begin{split}
 \lambda_{1} &= 0 \\
\lambda_{2,3} &= .4558 \,\pm \,\mathrm{j} 63.226 \\
\lambda_{4,5}&= -.006 \,\pm \, \mathrm{j} 7.246 \\
\lambda_{6} &= 0 \\
\end{split}
\end{align}

which are simply the algebraic negative of the eigenvalues for stable spin. Note that,now, \textit{the coupled frequency has a positive real part} which means it is unstable. Thus, when the rattleback is spun in the unstable direction, the coupled frequency, because of the large positive real part, is unstable and causes oscillations in pitch to grow and this high frequency "rattle" can be observed in any reversing rattleback. Once spin crosses zero, the eigenvalues change sign and the coupled frequency becomes stable and starts decaying as energy is transferred to the spinning and ultimately it dies down completely. 

\subsubsection{Rattleback - II}

To further investigate the validity of the linearization, simulations were run for a rattleback with different parameters(SI units) as summarized in Tab. \ref{hugeratt}.  These parameters were purposefully made to be large. 
\begin{center}
 \captionof{table}{SIMULATION PARAMETERS FOR RATTLEBACK -II}
\begin{tabular}{ c | c  }
  \hline                       
  $M$ & $ 4.5 \,\, \mathrm{kg} $  \\
  $a_1$ & $ .845 \,\, \mathrm{m} $\\
  $a_2$ & $ .268 \,\, \mathrm{m}$  \\
  $a_3$ &  $ .259 \,\, \mathrm{m}$  \\
      \hline  
    \end{tabular} \label{hugeratt}
\end{center}

Using Eqn. \ref{eigenequation} we can obtain the eigenvalues of the Jacobian for this rattleback as 

\begin{align}
\begin{split}
\lambda_{1,2} &= \pm \,\mathrm{j}10.62 \\
&\implies f_1 = \frac{\mathrm{abs}(\lambda_{1,2}) }{2 \pi} = 1.69 \, \, \mathrm{Hz}\\
\lambda_{3,4} &= \pm \, \mathrm{j}1.203 \\
&\implies f_2 = \frac{\mathrm{abs}(\lambda_{1,2}) }{2 \pi} = .206 \, \, \mathrm{Hz}\\
\end{split}
\end{align}

\begin{figure}[htp]
\centering
\includegraphics[width=\linewidth ]{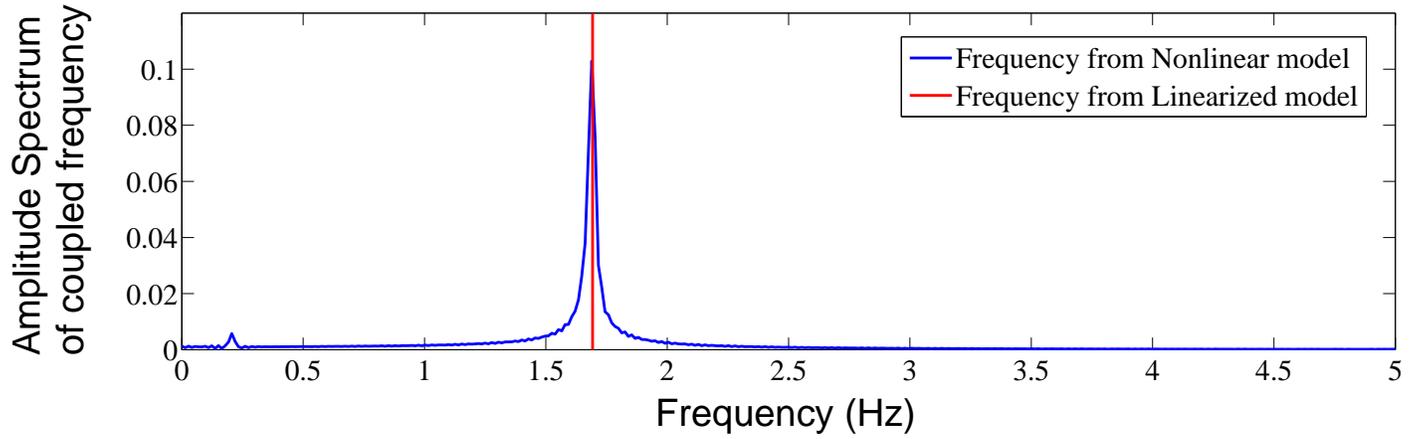}
\caption{ Amplitude spectrum of coupled frequency of a simulated rattleback with initial $\vect{\omega} =[0 \,\,0\,\, 0]^T \mathrm{rad/s}$ and $\vect{\hat{u}} = [.02 \,\,.02\,\, -.9996]^T$  }
\label{hugeratt1} 
\end{figure}

\begin{figure}[htp]
\centering
\includegraphics[width=\linewidth]{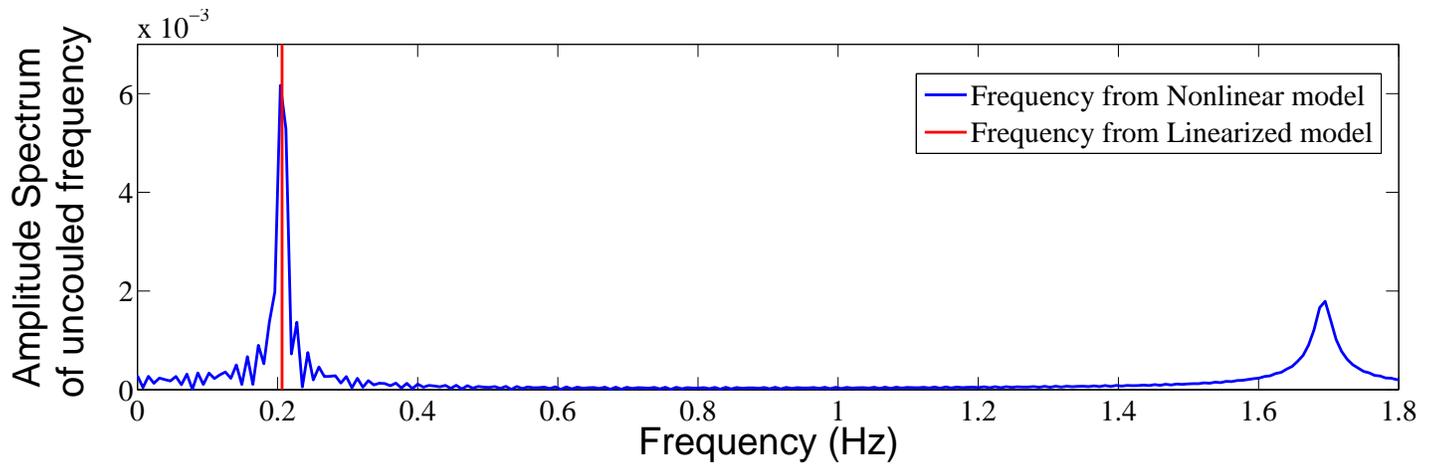}
\caption{ Amplitude spectrum of uncoupled frequency of a simulated rattleback with initial $\vect{\omega} =[0 \,\,0\,\, 0]^T \mathrm{rad/s}$ and $\vect{\hat{u}} = [.02 \,\,.02\,\, -.9996]^T$  }
\label{hugeratt2}
\end{figure}

Shown in Fig. \ref{hugeratt1} and Fig. \ref{hugeratt2} are the Amplitude spectra of the angular velocity of the simulated rattleback for the coupled and uncoupled frequencies respectively. The red line is the predicted frequency value obtained from the Linearized model via Eqn.\ref{eigenequation}. It can be seen that the frequency coincides almost exactly with the predicted values.

\begin{figure}[htp]
\centering
\includegraphics[width=\linewidth]{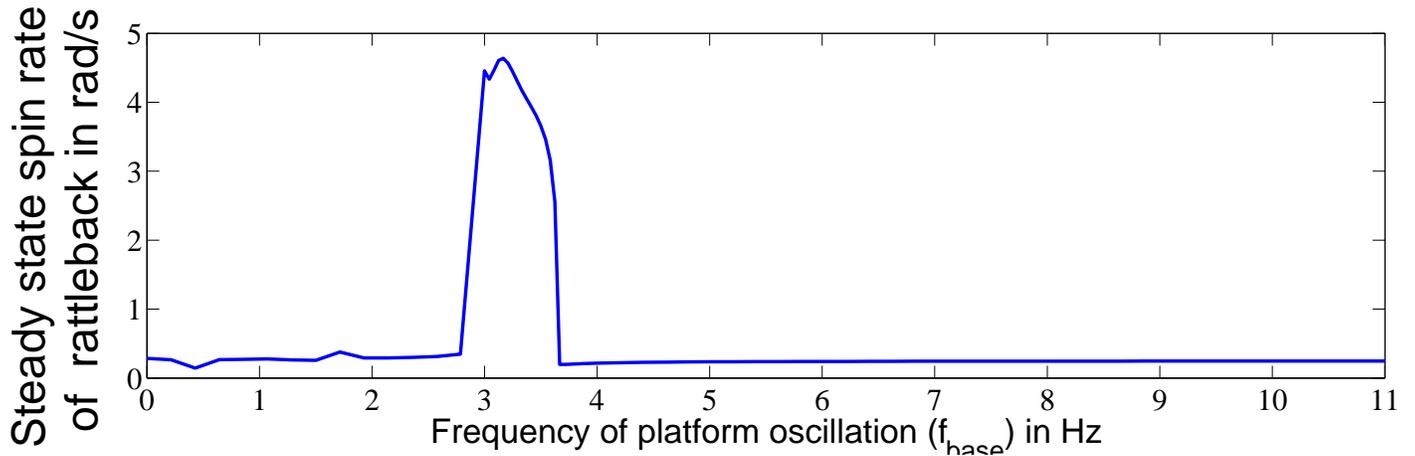}
\caption{ Plot of steady state spin rate of a simulated rattleback against frequency of platform oscillation with initial $\vect{\omega} =[0 \,\,0\,\, 0]^T \mathrm{rad/s}$ and $\vect{\hat{u}} = [.02 \,\,.02\,\, -.9996]^T$. Rattleback parameters in Tab. \ref{hugeratt} }
\label{spinresonance3}
\end{figure}

 Since the coupled frequency for this rattleback is $1.69 \, \, \mathrm{Hz}$, using Eqn. \ref{resonant}, we can expect to see a spin resonance at $f_{base } = 3.4 \,\, \mathrm{ Hz}$. Shown in Fig. \ref{spinresonance3} is a plot of steady state spin rate against frequency of base oscillations that shows clearly that this indeed is the case. 
 
\section{PROPOSED APPLICATIONS OF SPIN RESONANCE \label{appli}}

\begin{figure}[htb]
\centering
\includegraphics[width=1.15\linewidth]{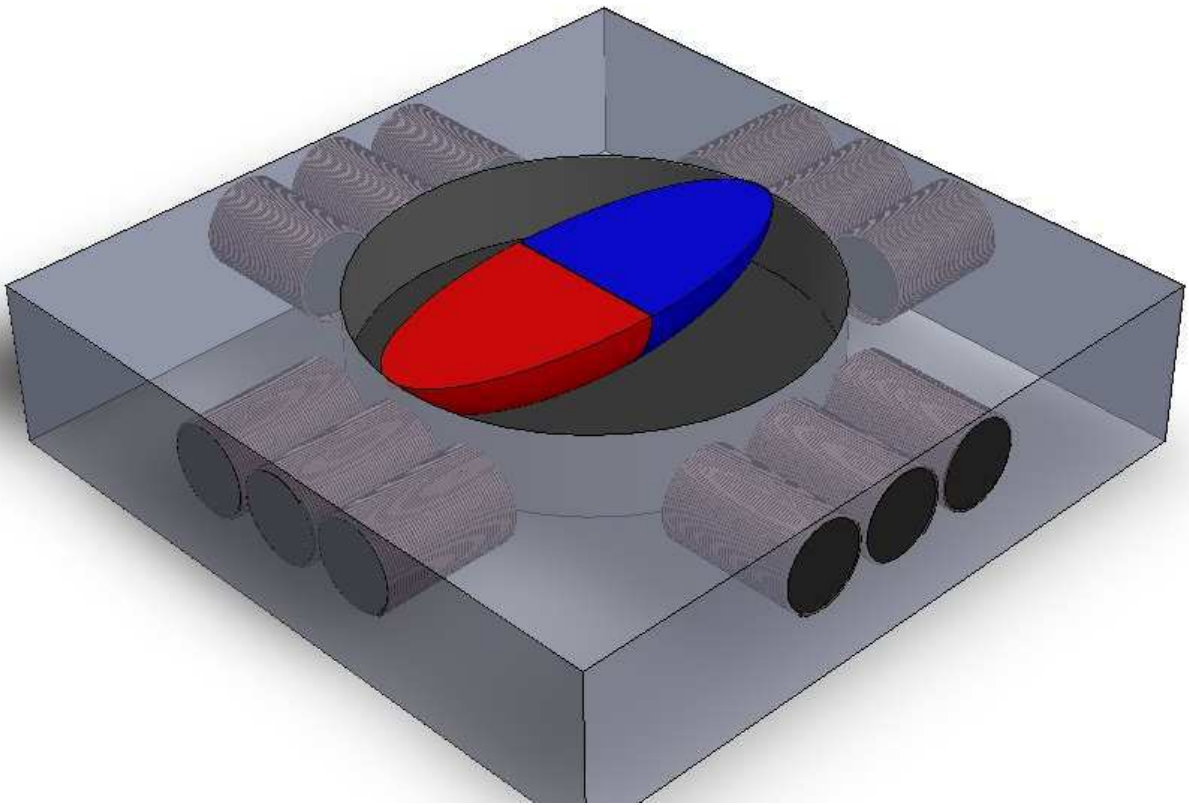} 
\caption{PROTOTYPE RATTLEBACK ENERGY HARVESTER}
\label{energyharvester}
\end{figure}

The conversion of vibrations(or linear translational motion) to rotations has been an important and desirable goal in many scenarios throughout the history of science and technology. Large scale examples include the piston mechanism and circular gears. In \cite{vibrot}, the authors present a simple device, which they refer to as "Vibrot",  consisting of a mass with 2 or more cantilever "legs" that make an angle with the vertical in such a way that placing it on a vertically  oscillating surface causes the device to rotate about the vertical axis; via a stick-slip mechanism.  In \cite{heckel2012}, Heckel \textit{et.al} present a circular ratchet with asymmetrical saw-like teeth filled with granular material that can achieve transduction of vibration into rotations. In \cite{norden2002}, Norde´n \textit{et.al} present a device similar to vibrot that is able to transduce vibrations along the vertical axis to unidirectional rotation about the same axis as a result of interaction between dry friction and inertia of the device, is analyzed. In \cite{liu2010}, Liu \textit{et.al} propose a design for an ultrasonic motor that achieves rotation by using Piezoelectric actuation to induce a traveling wave in a cylinder which then causes the attached rotor to rotate due to friction. 

Spin resonance can transduce vibrations to rotations and this transduction can be achieved very simply. Like many of the designs above, spin resonance is dependent on friction- since it is  friction that provides the torque that enables a rattleback to reverse spin or start spinning. However, the rattleback is fairly less complicated- it is a single component that can be easily manufactured which makes it ideal for applications for harvesting ambient energy. As a prototype for a energy harvester, consider a small rattleback fitted inside a disc shaped hole carved in a thick sheet and able to rotate with minimum dissipation as shown in Fig. \ref{energyharvester}. The sheet has coils of a ferromagnetic material embedded in it as shown and the rattleback is composed of a strongly magnetized material. If the sheet is subjected to oscillations in the neighborhood of the resonant frequency, then the magnetized rattleback will start spinning and it will cause small amounts of electricity to flow in the coils because of  electromagnetic induction   

\begin{figure}[htp]
\centering
\includegraphics[width=1.15\linewidth]{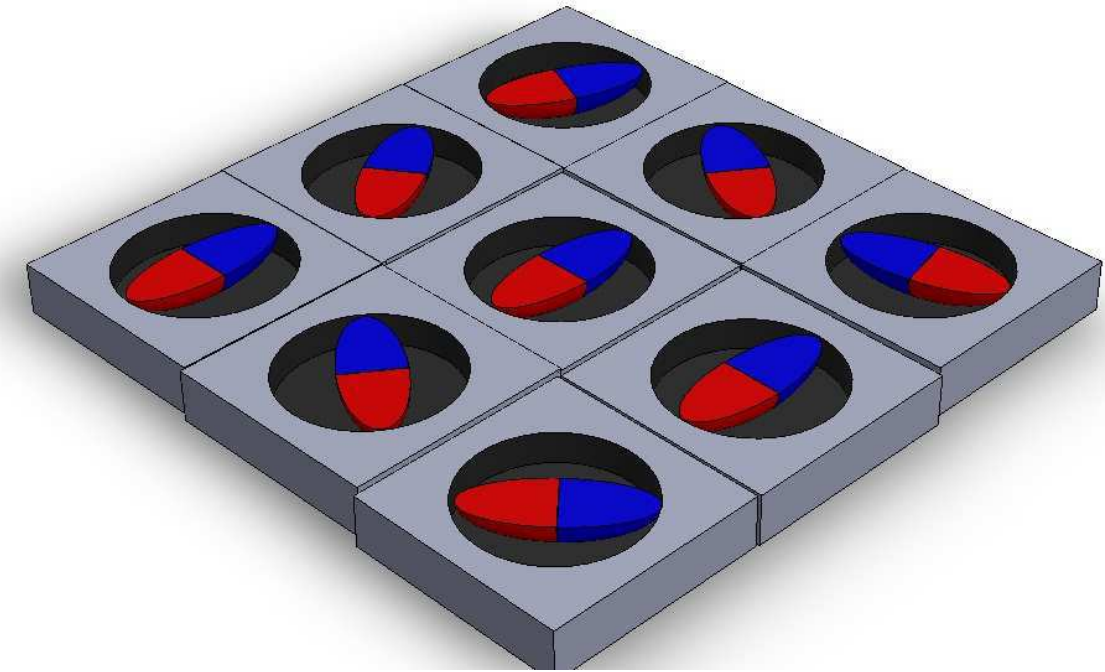} 
\caption{PROTOTYPE RATTLEBACK VIBRATION SENSOR}
\label{sensor}
\end{figure}

   Another prototype that the authors would like to propose is a design for a sensor. Consider the device in Fig. \ref{energyharvester} and we have many such rattlebacks arranged like an array, each with a different resonant frequency. Figure \ref{sensor} shows the arrangement. The coils are not shown for clarity. Since the resonant frequencies are different, it will be possible to estimate the frequency of base vibration by noting the rattlebacks that spin(and produce current). The accuracy of such an design will depend on the spread of the resonant frequencies and the number of rattlebacks used.    
   
 \section{ CONCLUSIONS AND FUTURE RESEARCH \label{conclu}}
 In this paper, the rattleback was modeled as a semi-ellipsoid and the governing equations  of motion were simulated to investigate the behavior when the rattleback is placed on an oscillating platform.  
 
  The response of rattleback was found to be composed of two fundamental frequencies that appeared irrespective of whether the rattleback was reversing spin or building spin from zero spin. Further, only the coupled frequency was found to be coupled to the spin and spin resonance was found to occur when the frequency of base oscillation was twice that of the coupled frequency. A linearized model of the dynamic equations was developed and analytical expressions that predict the fundamental frequencies of any given rattleback with reasonable accuracy were derived. A simple model for energy harvesting consisting of a magnetized rattleback contained within ferromagnetic walls (see Fig. \ref{energyharvester}) was also proposed. 
  
  This opens up scope for further research. An analysis can be done on the magnitude of electric power that can be extracted from vibration using the configuration in Fig. \ref{energyharvester}. An investigation can also be conducted on the efficacy and ability  of the rototypical Vibration Sensor(See Fig. \ref{sensor}) suggested in this paper to accurately resolve and estimate the frequency of ambient vibrations. The dependence between the coupled frequency and the spin motion which was found to have higher order nonlinear dynamics in this analysis can reveal ways of increasing the extent of coupling and thereby maximizing the energy that can be extracted from the vibrating platform and channeled into spinning kinetic energy. Further, only sinusoidal oscillations were considered in this analysis. Future research can investigate the effect of using non-sinusoidal oscillations on the spin resonance. 
  
\section{ACKNOWLEDGMENTS}

This research was partially supported by the Sustainability Program from the State University of New York Research Foundation and the authors would like to extend their gratitude to the same.  
   
\bibliographystyle{unsrt}

\end{document}